\documentclass[smallextended]{svjour3}
\smartqed 
\usepackage{amssymb,amsmath,graphicx,color}
\usepackage{amsfonts}
\usepackage{epsfig,latexsym,graphicx}

\begin{document}

\title{\bf{Asymptotic Properties of Minimum $S$-Divergence Estimator for Discrete Models}
\thanks{This is part of the Ph.D. research work of the author
 which is ongoing under the supervision of Prof. Ayanendranath Basu at the Indian Statistical Institute}
}
\titlerunning{Asymptotic Properties of MSDE}        
\author{Abhik Ghosh}
\institute{Abhik Ghosh \at
              Indian Statistical Institute \\
              203 B. T. Road, Kolkata 700 108, India\\
              \email{abhianik@gmail.com}           
}
\date{Received: date / Accepted: date}
\maketitle

\begin{abstract}
Robust inference based on the minimization of statistical divergences has proved 
to be a useful alternative to the classical techniques based on maximum likelihood and related methods. 
Recently Ghosh et al.~(2013) proposed a general class of divergence measures, namely the $S$-Divergence Family 
and discussed its usefulness in robust parametric estimation through some numerical illustrations. 
In this present paper, we develop the asymptotic properties of the proposed minimum $S$-Divergence 
estimators under discrete models.
\end{abstract}

\section{Introduction}
 
Density-based minimum distance methods provide attractive alternatives to 
likelihood based methods in parametric inference. 
Often these estimators combine strong robustness properties with full asymptotic efficiency. 
The estimators based on the family of power divergences (Cressie and Read, 1984) is one such example. 
Consider the class $\mathcal{G}$ of all probability density functions on the $\sigma$-field $(\Omega, \mathcal{A})$.
Usually, in practice $\Omega=\mathbb{R}^p$ and $\mathcal{A}$ is the corresponding Borel $\sigma$-field. 
The power divergence measure between two densities $g$ and $f$ in $\mathcal{G}$, indexed by a parameter 
$\lambda \in \mathbb{R}$, is defined as 
$$
{\rm PD}_\lambda( g, f) = \frac{1}{\lambda(\lambda + 1)} \int  g [ ( g/f)^\lambda - 1 ].
$$
Here, the integral is taken over the whole sample space $\Omega$ and 
the same is to be understood in the rest of the paper also, unless mentioned otherwise.  
For values of $\lambda = 1, 0,-1/2,-1$ and $ -2$ the family generates the Pearson's 
chi-square (PCS), the likelihood disparity (LD), the twice squared Hellinger distance (HD), 
the Kullback-Leibler divergence (KLD) and the Neyman's chi-square (NCS) respectively. 
The family is a subclass of the larger family of $\phi$-divergences (Csisz\'{a}r, 1963) or disparities. 
The minimum disparity estimator of $\theta$ under the model 
$\mathcal{F}=\{F_\theta : \theta \in \Theta \subseteq \mathbb{R}^p\}$ is the minimizer of the 
divergence between $\hat{g}$ (a nonparametric estimate  of the true density $g$) and the model density $f_\theta$. 
All minimum distance estimators based on disparities have the 
same influence function as that of the maximum likelihood estimator (MLE) at the model and 
hence have the same asymptotic model efficiency.

The evaluation of a minimum distance estimator based on disparities requires kernel density estimation, 
and hence inherits all the complications of the latter method. Basu et al.~(1998) developed 
a class of density-based divergence measures called the density power divergence (DPD) that 
produces robust parameter estimates but needs no nonparametric smoothing. 
The DPD measure between two densities $g$ and $f$ in $\mathcal{G}$ is defined, 
depending on a nonnegative parameter $\alpha$, as
\begin{eqnarray}
d_\alpha(g,f) &=& \int  f^{1+\alpha} - \frac{1+\alpha}{\alpha} \int f^\alpha g + 
\frac{1}{\alpha} \int g^{1+\alpha},  ~~~~ \mbox{ for } \alpha > 0, \nonumber \\
\mbox{and} ~~~~~~~~~~~~~~~~~~ && \nonumber \\
 d_0(g,f) &=& \lim_{\alpha \rightarrow 0} d_\alpha(g,f) = \int g \log(g/f).
\end{eqnarray}
The parameter $\alpha$ provides a smooth bridge between the likelihood disparity ($\alpha = 0$) and 
the $L_2$-divergence ($\alpha = 1$); it also controls the trade-off between the robustness and 
efficiency with larger $\alpha$ being associated with greater robustness but reduced efficiency. 
Both the PD and DPD families provide outlier down weighting using powers of model densities.

Combining the concepts of the power divergence and the density power divergence, 
Ghosh et al.~(2013) developed a two parameter family of density-based divergences, named as ``$S$-Divergence", 
that connects the whole of the Cressie-Read family of power divergence smoothly to 
the $L_2$-divergence at the other end. This family contains both the PD and
DPD families as special cases. Through various numerical examples, they illustrate that the minimum divergence
estimators based on the $S$-Divergence are also extremely robust and  are also competitive 
in terms efficiency for most of the members of this family.

In this present article, we will develop the theoretical properties of the minimum $S$-Divergence estimators. 
For simplicity, here we consider only the set up for the discrete model so that the true data generating 
probability mass functions can be estimated non-parametrically by just the relative frequencies of 
the observed sample --- we do not need to consider any nonparametric smoothing. 
We will prove the consistency and asymptotic normality of the minimum $S$-Divergence estimators. 
We will introduce the $S$-divergence and the minimum $S$-divergence estimator in Section \ref{SEC:S-div} 
and \ref{SEC:MSDE} respectively. Then Section \ref{SEC:asymptotics} will contain the 
asymptotic properties of the  minimum $S$-Divergence estimators. 
We will present the application of the minimum $S$-divergence estimator in some interesting real data examples 
in Section \ref{SEC:example} and relate the findings with the theoretical results in Section \ref{SEC:choice_par}, 
that leads to some indication on the choice of the tuning parameters. 
Finally we conclude the paper by  an overall conclusion in Section \ref{SEC:conclusion}. 
Throughout the rest of the paper, we will use the term ``density" for the probability mass functions also.


\section{The $S$-Divergence Family}\label{SEC:S-div}
 
It is well-known that the estimating equation for the minimum 
density power divergence represents an 
interesting density power down-weighting, and hence robustification of the usual likelihood score 
equation (Basu et al., 1998). The usual estimating equations for the MLE can be recovered from that estimating equation by the choice
$\alpha = 0$. Within the given range of $0 \leq \alpha \leq 1$, $\alpha =1 $ will lead to the maximum
down-weighting for the score functions of the surprising observations corresponding to the $L_2$ 
divergence ; on the other extreme, the score functions will be subjected to no down-weighting at 
all for $ \alpha = 0 $ corresponding to the Kullback-Leibler divergence (Kullback and Leibler, 1951). 
Intermediate values of $\alpha$ provide a smooth bridge between these two estimating equations, and
 the degree of down weighting increases with increasing $\alpha$.

Noting that the Kullback-Leibler divergence is a particular case of Cressie-Read family of 
power divergence corresponding to $\lambda = 0$, we see that the density power divergence gives us a smooth 
bridge between one particular member of the Cressie-Read family and the $L_2$ divergence with increasing robustness. 
Ghosh et al.~(2013) constructed a family of divergences which connect, 
in a similar fashion, other members of the PD family with the $L_2$-divergence. 
That larger super-family, named as the $S$-Divergence Family, is defined as

\begin{eqnarray}
		S_{(\alpha, \lambda)}(g,f) &=& \frac{1}{1+\lambda (1-\alpha)} ~ \int ~  
\left[\left( f^{1+\alpha} - g^{1+\alpha} \right) \right. \nonumber \\ 
&& ~~ - \left. \frac{(1+\alpha)}{(\alpha-\lambda (1-\alpha))}
 g^{1 + \lambda (1-\alpha)} \left(f^{\alpha-\lambda (1-\alpha)} - g^{\alpha-\lambda (1-\alpha)}\right)\right]  \nonumber
\\ \nonumber \\
 &=&  \frac{1}{A} ~ \int ~ f^{1+\alpha}  -   \frac{1+\alpha}{A B} ~ \int ~~ f^{B} g^{A}  
+ \frac{1}{B} ~ \int ~~ g^{1+\alpha},
\label{EQ:S_div_defn} \\ \nonumber 
\end{eqnarray}
where $A = 1+\lambda (1-\alpha)$ and   $B = \alpha - \lambda (1-\alpha)$.

Note that, $ A+B=1+\alpha $. Also the above form of divergence family is defined for those 
$\alpha$ and $\lambda$ for which $A \ne 0$ and $B \ne 0$.
If $A=0$ then the corresponding divergence measure is defined as the continuous limit of (\ref{EQ:S_div_defn}) as 
$A \rightarrow 0$ and is given by
\begin{eqnarray}
    S_{(\alpha,\lambda : A = 0)}(g,f) &=& \lim_{A \rightarrow 0} ~ S_{(\alpha, \lambda)}(g,f) \nonumber \\
     &=&  \int f^{1+\alpha} \log(\frac{f}{g}) - \int \frac{(f^{1+\alpha} - g^{1+\alpha})}{{1+\alpha}}.
\label{EQ:S_div_defn_A0}
\end{eqnarray}
Similarly, if $B=0$ then the divergence measure is defined to be
\begin{eqnarray}
    S_{(\alpha,\lambda : B = 0)}(g,f) &=& \lim_{B \rightarrow 0} ~ S_{(\alpha, \lambda)}(g,f) \nonumber \\
         &=&  \int g^{1+\alpha} log(\frac{g}{f}) - \int \frac{(g^{1+\alpha} - f^{1+\alpha})}{{1+\alpha}}.
\label{EQ:S_div_defn_B0}
\end{eqnarray}

Note that for $\alpha = 0$, the class of $S$-divergences reduces to the PD family  with parameter $\lambda$; 
for $\alpha =1$, $S_{1, \lambda}$ equals the $L_2$ divergence irrespective of the value of $\lambda$. 
On the other hand, $\lambda = 0$ generates the DPD family as a function of $\alpha$.
In Ghosh et al.~(2013a), it was shown that  
the above $S$-divergence  family defined in (\ref{EQ:S_div_defn}), (\ref{EQ:S_div_defn_A0}) 
and (\ref{EQ:S_div_defn_B0}) indeed represent a family of genuine statistical divergence measures in the sense
that $S_{(\alpha, \lambda)}(g,f) \ge 0$ for densities $g, f $ and all $\alpha \ge 0$, $\lambda \in \mathbb{R}$,
and $S_{(\alpha, \lambda)}(g, f)$ is equal to zero if and only if $g=f$ identically.

\section{The Minimum $S$-Divergence Estimators}\label{SEC:MSDE}

Let us now consider the discrete set-up for parametric estimation. 
Let $X_1$, $\cdots$, $X_n$ denotes $n$ independent and identically
distributed observations from the true distribution $G$ having a probability density function $g$ 
with respect to some counting measure. Without loss of generality, we will assume that the support 
of $g$ is $\chi = \{0, 1, 2, \cdots \}$. Let us denote the relative frequency at $x$ obtained from data by 
$r_n(x) = \frac{1}{n} \sum_{i=1}^n I(X_i = x)$, where $I(A)$ denotes the indicator function of the event $A$. 
We model the true data generating distribution $G$ by the parametric model family 
$\mathcal{F}=\{F_\theta : \theta \in \Theta \subseteq \mathbb{R}^p\}$. 
We will assume that both $G$ and $\mathcal F$ belong to $\mathcal{G}$, 
the (convex) class of all distributions having densities with respect to the counting measure (or the 
appropriate dominating measure in other cases). We are interested in the estimation of the parameter $\theta$.

Note that, the minimum $S$-divergence estimator has to be obtained by minimizing the $S$-divergence  
measure between the data and the model distribution. 
However, in the discrete set-up, both the data-generating true distribution and the model distribution are 
characterized by the probability vectors  $\mathbf{r}_n = ( r_n(0), r_n(1), \cdots )^T$ and 
$\mathbf{f}_\theta = ( f_\theta(0), f_\theta(1), \cdots )^T$ respectively. Thus in this case, 
the minimum $S$-divergence  estimator of $\theta$ can be obtained by just minimizing 
$S_{(\alpha, \lambda)}(\mathbf{r}_n,\mathbf{f}_\theta)$, the $S$-divergence  measure between
 $\mathbf{r}_n$ and $\mathbf{f}_\theta$, with respect to $\theta$. 
The estimating equation is then given by
\begin{eqnarray}
&& \nabla S_{(\alpha, \lambda)}(\mathbf{r}_n,\mathbf{f}_\theta) = 0, \nonumber\\
\mbox{or,   } && \nabla \left[ \frac{1}{A} ~ \sum_{x=0}^\infty f_{\theta}^{1+\alpha}(x)  
-   \frac{1+\alpha}{A B} ~ \sum_{x=0}^\infty f_{\theta}^{B}(x) r_n^{A}(x)  
+ \frac{1}{B} ~ \sum_{x=0}^\infty r_n^{1+\alpha}(x) \right] =0, \nonumber\\
\mbox{or,   } &&  \frac{1+\alpha}{A} \sum_{x=0}^\infty f_{\theta}^{1+\alpha}(x) u_{\theta}(x) 
- \frac{1+\alpha}{A} \sum_{x=0}^\infty f_{\theta}^{B}(x) r_n^{A}(x) u_{\theta}(x) = 0, \\
\mbox{or,   } && \sum_{x=0}^\infty K(\delta(x))f_{\theta}^{1+\alpha}(x) u_{\theta}(x) = 0,
\label{EQ:S-est_eqn_discrete}
\end{eqnarray}
where $\delta(x)= \delta_n(x) = \frac{r_n(x)}{f_{\theta}(x)} - 1$, $ K(\delta) = \frac{(\delta+1)^A - 1}{A}$ 
and $u_\theta(x)=\nabla \ln f_\theta(x)$ is the likelihood score function. 
Note that, $\nabla$ represents the derivative with respect to $\theta$ and 
we will denote its $i^{\rm th}$ component by $\nabla_i$.

\section{Asymptotic properties of the Minimum $S$-Divergence Estimators}
 \label{SEC:asymptotics}

	Now we will derive the asymptotic properties of the minimum $S$-divergence estimator 
under the discrete set-up as mentioned above.  Note that, in order to obtain the 
minimum $S$-divergence estimator under discrete set-up, we need to minimize 
$S_{(\alpha, \lambda)}(\mathbf{r}_n,\mathbf{f}_\theta)$ 
over $\theta$ which is equivalent to minimizing  $H_n(\theta)$ with respect to $\theta$ where
\begin{eqnarray}
    H_n(\theta) = \frac{1}{1+\alpha} \left[ \frac{1}{A} \sum_x f_{\theta}^{1+\alpha}(x) 
- \frac{1+\alpha}{AB} \sum_x f_{\theta}^B(x) r_n^A(x) \right].
\end{eqnarray}
Now,
\begin{eqnarray}
    \nabla H_n(\theta) &=& \frac{1}{A} \left[ \sum_x f_{\theta}^{1+\alpha}(x)u_{\theta}(x) 
-  \sum_x f_{\theta}^B(x) u_{\theta}(x) r_n^A(x) \right]  \nonumber \\
  &=&  - \sum_x K(\delta_n(x)) f_{\theta}^{1+\alpha}(x) u_{\theta}(x),
\end{eqnarray}
where $\delta_n(x) = \frac{r_n(x)}{f_{\theta}(x)} - 1$. Thus the estimating equation is exactly the same as given in 
Equation (\ref{EQ:S-est_eqn_discrete}). Let $\theta^g$ denotes the ``best fitting parameter" 
under the true density $g$, obtained by minimizing $S_{(\alpha, \lambda)}(g,\mathbf{f}_\theta)$
over the parameter space $\theta \in \Theta$. Define 
\begin{eqnarray}
J_g &=& J_\alpha(g) = E_g \left[ u_{\theta^g}(X)u_{\theta^g}^T(X) K'(\delta_g^g(X)) f_{\theta^g}^\alpha(X) \right]
\nonumber \\
&& ~~~~~~~~~~~~~~~~~~~~~~~~~~ - \sum_x  K(\delta_g^g(X)) \nabla^2 f_{\theta^g}(x), \\
V_g &=& V_\alpha(g) = V_g \left[ K'(\delta_g^g(X)) f_{\theta^g}^\alpha(X) u_{\theta^g}(X) \right],
\end{eqnarray}
where $X$ denotes a random variable having density $g$, 
$\delta_g(x) = \frac{g(x)}{f_\theta(x)} -1$, $\delta_g^g(x) = \frac{g(x)}{f_{\theta^g}(x)} -1$, 
$K'(\cdot)$ is the derivative of $K(\cdot)$ with respect to its argument and 
$\nabla^2$ represent the second order derivative with respect to $\theta$.
We will prove the asymptotic properties of the minimum $S$-divergence estimator under the following assumptions:

\begin{enumerate}
\item[(SA1)] The model family $\mathcal{F}$ is identifiable, i.e.,  
for any two $F_{\theta_1}$ and $F_{\theta_2}$ in the model family $\mathcal{F}$, 
    $$
    F_{\theta_1} = F_{\theta_2}  \Rightarrow \theta_1 = \theta_2.
    $$
\item[(SA2)] The probability density function $f_\theta$ of the model distribution have common support 
so that the set $\chi = \{ x : f_{\theta}(x) >0 \}$ is independent of $\theta$.
			Also the true distribution $g$ is compatible with the model family.
\item[(SA3)] There exists an open subset $\omega \subset \Theta$
 for which the best fitting parameter $\theta^g$ is an interior point and for almost all x, the density 
$f_{\theta}(x)$ admits all the third derivatives of the type $\nabla_{jkl} f_{\theta}(x)$   
$\forall \theta \in \omega$. Here, $\nabla_{jkl}$ denotes the $(j, k, l)^{\rm th}$ element of $\nabla^3$, 
the third order derivative with respect to $\theta$
\item[(SA4)] The matrix $J_g$ is positive definite.
\item[(SA5)] The quantities  $\sum_x g^{1/2}(x) f_{\theta}^{\alpha}(x) |u_{j\theta}(x)|$, 
$\sum_x g^{1/2}(x) f_{\theta}^{\alpha}(x) |u_{j\theta}(x)| |u_{k\theta}(x)|$ and 
 $\sum_x g^{1/2}(x) f_{\theta}^{\alpha}(x) |u_{jk\theta}(x)|$  are bounded $\forall  j,k$ 
and $\forall \theta \in \omega$.  \\
Here, $u_{j\theta}(x)$ denotes the $j^{\rm th}$ element of $u_{\theta}(x)$ and  $u_{jk\theta}(x)$ denotes 
the $(j, k)^{\rm th}$ element of $\nabla^2 \ln f_{\theta}(x)$.
\item[(SA6)] For almost all $x$, there exists functions $M_{jkl}(x)$, $M_{jk,l}(x)$, $M_{j,k,l}(x)$ 
that dominate, in absolute value,  $f_{\theta}^{\alpha}(x) u_{jkl\theta}(x)$, 
$f_{\theta}^{\alpha}(x) u_{jk\theta}(x) u_{l\theta}(x)$ and \\
$f_{\theta}^{\alpha}(x) u_{j\theta}(x) u_{k\theta}(x) u_{l\theta}(x)$ 
respectively $ \forall j, k, l$ and that are uniformly bounded in expectation with respect to $g$ and 
$f_{\theta} ~~ \forall \theta \in \omega$. \\
Here, $u_{jkl\theta}(x)$ denotes the $(j, k, l)^{\rm th}$ element of $\nabla^3 \ln f_{\theta}(x)$.
\item[(SA7)] The function $\left( \frac{g(x)}{f_{\theta}(x)} \right)^{A-1}$ is uniformly bounded (by, say, $C$) 
$\forall \theta \in \omega$.
\\
\end{enumerate}

	To prove the consistency and asymptotic normality of the minimum $S$-divergence estimator, we will, now on,
 assume that the above $7$ conditions hold. We will first consider some Lemmas.

\begin{lemma}
Define $ \eta_n(x) = \sqrt n \left(\sqrt{\delta_n(x)}-\sqrt{\delta_g(x)}\right)^2$. 
For any $k \in [0,2]$ and any $x \in \chi$, we have
\begin{enumerate}
    \item $E_g[\eta_n(x)^k] \le n^{\frac{k}{2}}E_g[|\delta_n(x) - \delta_g(x)|]^k \le 
\left[ \frac{g(x)(1-g(x))}{f_{\theta}^2(x)}\right]^{\frac{k}{2}}$. \\
	 \item   $E_g[|\delta_n(x) - \delta_g(x)|] \le \frac{2g(x)(1-g(x))}{f_{\theta}(x)}$.
\end{enumerate}
\label{LEM:lemma_3.1}
\end{lemma}
\noindent
\textbf{Proof :} 
The proof uses the same argument as in Lemma 2.13 of Basu et al.~(2013, page 56). 
For $a,b \ge 0$, we have the inequality $(\sqrt a - \sqrt b)^2 \le |a-b|$. So we get
\begin{eqnarray}
E_g[\eta_n(x)^k] &=& n^{\frac{k}{2}} E_g\left[\left(\sqrt{\delta_n(x)}-\sqrt{\delta_g(x)}\right)^{2k}\right]
\nonumber\\ 
&=& n^{\frac{k}{2}} E_g\left[\left(\sqrt{\delta_n(x)}-\sqrt{\delta_g(x)}\right)^{2}\right]^k\nonumber \\
& \le & n^{\frac{k}{2}} E_g[|\delta_n(x) - \delta_g(x)|]^k. \nonumber 
\end{eqnarray}

	For the next part see that, under $g$, $ n r_n(x) \sim Binomial( n, g(x) ) ~~ \forall x$. 
Now, for any $k \in [0,2]$, we get by the Lyapounov's inequality that 
\begin{eqnarray}
E_g[|\delta_n(x) - \delta_g(x)|]^k &\le & \left[ E_g(\delta_n(x) - \delta_g(x))^2 \right]^{\frac{k}{2}} 
\nonumber \\	
&=& \frac{1}{f_{\theta}^k(x)} \left[ E_g(r_n(x) - g(x))^2 \right]^{\frac{k}{2}} \nonumber \\		
&=& \frac{1}{f_{\theta}^k(x)} \left[ \frac{g(x)(1-g(x))}{n}\right]^{\frac{k}{2}}. \nonumber \\ 
\nonumber
\end{eqnarray}

	For the second part, note that
\begin{eqnarray}
E_g[|\delta_n(x) - \delta_g(x)|] &=& \frac{1}{f_{\theta}^k(x)} \left[ E_g|r_n(x) - g(x)| \right]^{\frac{k}{2}} 
\nonumber \\		
&\le& \frac{2g(x)(1-g(x))}{f_{\theta}(x)}, \nonumber 
\end{eqnarray}
where the last inequality follows from the result about the mean-deviation of a Binomial random variable.
\hfill{$\square$}

\begin{lemma}
	$E_g[\eta_n(x)^k] \rightarrow 0$, as $n \rightarrow \infty$,  for $k \in [0,2)$  and $x \in \chi$.
	\label{LEM:lemma_3.2}
\end{lemma}
\noindent
\textbf{Proof :}
This follows from Theorem 4.5.2 of Chung (1974) by noting that 
$n^{1/4} (r_n^{1/2}(x) - g^{1/2}(x)) \rightarrow 0$ with probability one for each $x \in \chi$ and 
by the Lemma \ref{LEM:lemma_3.1}(1), $\displaystyle \sup_n E_g[\eta_n^k(x)] $ is bounded.
\hfill{$\square$}

Let us now define, 
\begin{eqnarray}
a_n(x) &=& K(\delta_n(x)) - K(\delta_g(x)), \nonumber \\
b_n(x) &=& (\delta_n(x)-\delta_g(x))K'(\delta_g(x)), \nonumber \\
\mbox{and} ~~~~	\tau_n(x) &=& \sqrt n |a_n(x) - b_n(x)|.\nonumber
\end{eqnarray}
 We will need the limiting distributions of \\ \begin{center}
$S_{1n} = \sqrt n \displaystyle\sum_x a_n(x)f_{\theta}^{1+\alpha}(x)u_{\theta}(x)$ and 
$S_{2n} = \sqrt n \displaystyle\sum_x b_n(x)f_{\theta}^{1+\alpha}(x)u_{\theta}(x)$.
\end{center}
Next two Lemmas will help us to derive those distributions.

\begin{lemma}
	Assume condition (SA5). Then, 
	$$E_g|S_{1n}-S_{2n}| \rightarrow 0,  ~~~ \mbox{as  } ~ n \rightarrow \infty,$$ 
and hence 
$$S_{1n}-S_{2n} \displaystyle\mathop{\rightarrow}^\mathcal{P}  0, ~~~ \mbox{as  } ~ n \rightarrow \infty.$$ 
\label{LEM:lemma_3.3}
\end{lemma}
\noindent
\textbf{Proof :}
	By Lemma 2.15 of Basu et al.~(2011) [or, Lindsay (1994), Lemma 25], there exists 
some positive constant $\beta$ such that 
$$
\tau_n(x) \le \beta \sqrt n \left(\sqrt{\delta_n(x)}-\sqrt{\delta_g(x)}\right)^2 = \beta \eta_n(x).
$$ 
Also, by Lemma \ref{LEM:lemma_3.1}, $ E_g[\tau_n(x)] \le \beta \frac{g^{1/2}(x)}{f_{\theta}(x)}$.\\
And by Lemma \ref{LEM:lemma_3.2}, $E_g[\tau_n(x)] = \beta E_g[\eta_n(x)] \rightarrow 0$ as $n \rightarrow \infty$. Thus we get, 
\begin{eqnarray}
    E_g|S_{1n}-S_{2n}| &\le& \sum_x E_g[\tau_n(x)] f_{\theta}^{1+\alpha}(x)|u_{\theta}(x)| \nonumber \\
                     &\le&  \beta \sum_x g^{1/2}(x) f_{\theta}^{\alpha}(x)|u_{\theta}(x)| \nonumber \\
							& < &   \infty ~~~~~~ \mbox{ (by assumption SA5)}. \nonumber \\ \nonumber
\end{eqnarray}
So, by Dominated Convergence Theorem (DCT),  $E_g|S_{1n}-S_{2n}| \rightarrow 0 $ as $n \rightarrow \infty$.\\
Hence, by Markov inequality, $S_{1n}-S_{2n} \displaystyle\mathop{\rightarrow}^\mathcal{P}  0 $ as $n \rightarrow \infty$.
\hfill{$\square$}

\begin{lemma}
	Suppose $V_g $ is finite. Then under $g$, $$ S_{1n} \mathop{\rightarrow}^\mathcal{D}  N(0, V_g).$$
	\label{LEM:lemma_3.4}
\end{lemma}
\noindent
\textbf{Proof :}
Note that, by the previous Lemma \ref{LEM:lemma_3.3}, 
the asymptotic distribution of $S_{1n}$ and $S_{2n}$ are the same. 
Now, we have
\begin{eqnarray}
S_{2n} &=& \sqrt n \sum_x (\delta_n(x) - \delta_g(x)) K'(\delta_g(x)) f_{\theta}^{1+\alpha}(x) u_{\theta}(x) 
\nonumber \\
&=&  \sqrt n \sum_x (r_n(x) - g(x)) K'(\delta_g(x)) f_{\theta}^{\alpha}(x) u_{\theta}(x) \nonumber \\
&=&  \sqrt n \left( \frac{1}{n} \sum_{i=1}^n \left[ K'(\delta_g(X_i)) 
f_{\theta}^{\alpha}(X_i) u_{\theta}(X_i) - E_g\{ K'(\delta_g(X))f_{\theta}^{\alpha}(X)u_{\theta}(X)\} \right]\right) 
\nonumber \\
&\mathop{\rightarrow}^\mathcal{D} & Z \sim N( 0, V_g) ~~~ \mbox{[by Central Limit Theorem (CLT)]}. \nonumber
\end{eqnarray}
This completes the proof. \hfill{$\square$}

We will now consider the main theorem of this section about the consistency and asymptotic 
normality of the minimum $S$-divergence  estimator.\\

\begin{theorem}
	Under Assumptions (SA1)-(SA7), there exists a consistent sequence $\hat\theta_n$ of roots to 
the minimum $S$-divergence  estimating equation (\ref{EQ:S-est_eqn_discrete}).\\
	Also, the asymptotic distribution of $\sqrt n (\hat\theta_n - \theta^g)$ is $p-$dimensional
 normal with mean $0$ and variance $J_g^{-1}V_g J_g^{-1}$.\\
\end{theorem}
\noindent
\textbf{Proof of consistency:}	
Consider the behavior of $S_{(\alpha, \lambda)}(\mathbf{r}_n,\mathbf{f}_{\theta})$ on a sphere $Q_a$ 
which has  radius $a$ and center at $\theta^g$. We will show, for sufficiently small $a$, the
 probability tends to one that 
$$
S_{(\alpha, \lambda)}(\mathbf{r}_n,\mathbf{f}_{\theta}) > S_{(\alpha, \lambda)}(\mathbf{r}_n,\mathbf{f}_{\theta}) 
~~ \forall \theta ~~ \mbox{ on the surface of } Q_a,
$$
so that the $S$-divergence  has a local minimum with respect to $\theta$ in the interior of $Q_a$. At a local 
minimum, the estimating equations must be satisfied. Therefore, for any $a>0$ sufficiently small, 
the minimum $S$-divergence  estimating equation have a solution $\theta_n$ within $Q_a$ with probability 
tending to one as $n \rightarrow \infty$.

Now taking Taylor series expansion of $S_{(\alpha, \lambda)}(\mathbf{r}_n,\mathbf{f}_{\theta})$ 
about $\theta = \theta^g$, we get 
\begin{eqnarray}
&& S_{(\alpha, \lambda)}(\mathbf{r}_n,\mathbf{f}_{\theta}) - S_{(\alpha, \lambda)}(\mathbf{r}_n,\mathbf{f}_{\theta})
\nonumber \\
&& ~~~~ = - \sum_j(\theta_j-\theta_j^g)
\nabla_j S_{(\alpha, \lambda)}(\mathbf{r}_n,\mathbf{f}_{\theta})|_{\theta=\theta^g} 
\nonumber \\
&&  	~~~~~~~~~~~~~~~ - \frac{1}{2} 
\sum_{j,k}  (\theta_j - \theta_j^g)(\theta_k - \theta_k^g)
\nabla_{jk} S_{(\alpha, \lambda)}(\mathbf{r}_n,\mathbf{f}_{\theta})|_{\theta = \theta^g} 
\nonumber \\
&&  	~~~~~~~~~~~~~~~ -      \frac{1}{6} \sum_{j,k,l} (\theta_j - \theta_j^g)(\theta_k - \theta_k^g)
(\theta_l - \theta_l^g)\nabla_{jkl} S_{(\alpha, \lambda)}(\mathbf{r}_n,\mathbf{f}_{\theta})|_{\theta = \theta^*} 
\nonumber \\
&& ~~~~ = S_1 + S_2 + S_3, ~~~~~~~~ (say) \nonumber 
\end{eqnarray}
where $\theta^*$ lies between $\theta^g$ and $\theta$. We will now consider each terms one-by-one.

	For the Linear term $S_1$, we consider
\begin{eqnarray}
    \nabla_j S_{(\alpha, \lambda)}(\mathbf{r}_n,\mathbf{f}_{\theta})|_{\theta = \theta^g} = -(1+\alpha)~ 
\sum_x K(\delta_n^g(x))f_{\theta^g}^{1+\alpha}(x)u_{j\theta^g}(x),
\end{eqnarray}
where $\delta_n^g(x)$ is the $\delta_n(x)$ evaluated at $\theta = \theta^g$. 
We will now show that 
\begin{eqnarray}
    \sum_x K(\delta_n^g(x))f_{\theta^g}^{1+\alpha}(x)u_{j\theta^g}(x) 
\mathop{\rightarrow}^\mathcal{P} \sum_x K(\delta_g^g(x)) f_{\theta^g}^{1+\alpha}(x) u_{j\theta^g}(x),
\label{EQ:38}
\end{eqnarray}
as $n \rightarrow \infty$ and note that the right hand side of above is zero 
by definition of the Minimum $S$-divergence  estimator. Note that by assumption (SA7) and the fact that $r_n(x) \rightarrow g(x), $ almost surely (a.s.) by Strong Law of Large Number (SLLN), it follows that 
\begin{eqnarray}
	|K'(\delta)| = |\delta|^{A-1} < 2C = C_1,  ~~~~~~ \mbox{( say )}
	\label{EQ:39}
\end{eqnarray}
for any $\delta$ in between $\delta_n^g(x)$ and $\delta_g^g(x)$, uniformly in $x$. 
So, by using the one-term Taylor series expansion,
\begin{eqnarray}
 & &   |\sum_x K(\delta_n^g(x))f_{\theta^g}^{1+\alpha}(x)u_{j\theta^g}(x) - 
\sum_x K(\delta_g^g(x)) f_{\theta^g}^{1+\alpha}(x) u_{j\theta^g}(x) |  \nonumber \\
& &~~~~ \leq C_1 \sum_x |\delta_n^g(x) - \delta_g^g(x)| f_{\theta^g}^{1+\alpha}(x) |u_{j\theta^g}(x) |. \nonumber \\ \nonumber 
\end{eqnarray}
However, by Lemma \ref{LEM:lemma_3.1}(1), 
\begin{eqnarray}
E[|\delta_n^g(x) - \delta_g^g(x)|] \le \frac{\left[g(x)(1-g(x))\right]^{1/2}}{f_{\theta^g}(x) \sqrt n} \rightarrow 0,
 ~~~ ~~~ as ~ ~ n \rightarrow \infty.
\end{eqnarray} 
	and, by Lemma \ref{LEM:lemma_3.1}(2), we have 
\begin{eqnarray}
 & &  E[ C_1 \sum_x |\delta_n^g(x) - \delta_g^g(x)| f_{\theta^g}^{1+\alpha}(x) |u_{j\theta^g}(x) |] \nonumber \\
&& ~~~~~ \leq 2C_1 \sum_x g^{1/2}(x) f_{\theta^g}^{\alpha}(x) |u_{j\theta^g}(x) |  ~ < ~ \infty.     \\
&& ~~~~~~~~~~~~~~~~~~~~~~\mbox{[by assumption (A5)]} \nonumber
\end{eqnarray}
	Hence, by dominated convergence theorem (DCT), we get, 
\begin{eqnarray}
     E[|\sum_x K(\delta_n^g(x))f_{\theta^g}^{1+\alpha}(x)u_{j\theta^g}(x) - 
\sum_x K(\delta_g^g(x)) f_{\theta^g}^{1+\alpha}(x) u_{j\theta^g}(x) |] \rightarrow 0, 
\end{eqnarray}as  $n \rightarrow \infty$,
so that by Markov inequality we have the desired claim. Therefore, we have
\begin{eqnarray}
    \nabla_j S_{(\alpha, \lambda)}(\mathbf{r}_n,\mathbf{f}_{\theta})|_{\theta = \theta^g} \mathop{\rightarrow}^\mathcal{P} 0.
\end{eqnarray}
Thus, with probability tending to one, $|S_1| < p a^3$, where $p$ is the dimension 
of $\theta$ and $a$ is the radius of $Q_a$.\\


	Next we consider the quadratic term $S_2$. We have,
\begin{eqnarray}
    & &\nabla_{jk} S_{(\alpha, \lambda)}(r_n, f_\theta)|_{\theta=\theta^g} \nonumber \\
    && ~~~ = \nabla_k \left( -(1+\alpha) 
~ \sum_x K(\delta_n(x))f_{\theta}^{1+\alpha}(x)u_{j\theta}(x)|_{\theta=\theta^g} \right)\nonumber \\
    && ~~~ = -(1+\alpha) \left[ - \sum_x K'(\delta_n^g(x)) \delta_n^g(x) f_{\theta^g}^{1+\alpha}(x) 
u_{j\theta^g}(x) u_{k\theta^g}(x) \right. \nonumber \\
 & &  ~~~~~~~~~~~~~~~~~~~~~~ + \sum_x K(\delta_n^g(x))f_{\theta^g}^{1+\alpha}(x)u_{jk\theta^g}(x)  \nonumber \\
 & &  ~~~~~~~~~~~~~~~~~~~~~~ - \left. \sum_x K(\delta_n^g(x)) f_{\theta^g}^{1+\alpha}(x) u_{j\theta^g}(x) 
u_{k\theta^g}(x) \right]. \\ \nonumber  
\end{eqnarray}
	We will now show that 
\begin{eqnarray}
    &&  - \sum_x K'(\delta_n^g(x)) \delta_n^g(x) f_{\theta^g}^{1+\alpha}(x) u_{j\theta^g}(x) u_{k\theta^g}(x) \nonumber \\
    && ~~~~~~~~ \mathop{\rightarrow}^\mathcal{P}  - \sum_x K'(\delta_g^g(x)) \delta_g^g(x) f_{\theta^g}^{1+\alpha}(x) u_{j\theta^g}(x) u_{k\theta^g}(x).
 \label{EQ:45}
\end{eqnarray}
For note that as in (\ref{EQ:39}), we have
\begin{eqnarray}
    |K''(\delta)\delta| = |(A-1)||\delta|^{(A-1)} < C_2, ~~~ \mbox{(say)} \label{EQ:46}
\end{eqnarray}
for every $\delta$ lying in between $\delta_n^g(x)$ and $\delta_g^g(x)$, uniformly in $x$.
 So, by using the one-term Taylor series expansion, 
\begin{eqnarray}
   |K'(\delta_n^g)\delta_n^g - K'(\delta_g^g)\delta_g^g|  & \le & |\delta_n^g -\delta_g^g|
|K''(\delta_n^*)\delta_n^* + K'(\delta_n^*)| \nonumber \\
& \le &  |\delta_n^g -\delta_g^g| (C_2 +C_1). \nonumber \\  \nonumber 
\end{eqnarray}
Thus, we get 
\begin{eqnarray}
 & &      \left| \sum_x K'(\delta_n^g(x)) \delta_n^g(x) f_{\theta^g}^{1+\alpha}(x) u_{j\theta^g}(x) u_{k\theta^g}(x) 
 \right. \nonumber \\
 &&    ~~~~~~~~ \left.- \sum_x K'(\delta_g^g(x)) \delta_g^g(x) f_{\theta^g}^{1+\alpha}(x) u_{j\theta^g}(x) u_{k\theta^g}(x) \right| \nonumber \\
& & ~~~ \le (C_1 +C_2) \sum_x |\delta_n^g -\delta_g^g| f_{\theta^g}^{1+\alpha}(x) | u_{j\theta^g}(x) u_{k\theta^g}(x) |.
 \nonumber \\ \nonumber    
\end{eqnarray}
Since by assumption (SA5), we have $\sum_x g^{1/2}(x)f_{\theta^g}^{1+\alpha}(x) |u_{j\theta^g}(x) u_{k\theta^g}(x) | 
< \infty$, the desired result (\ref{EQ:45}) follows by the similar proof for proving (\ref{EQ:38}) above.
Similarly we also get that
\begin{eqnarray}
  \sum_x K(\delta_n^g(x)) f_{\theta^g}^{1+\alpha}(x) u_{jk\theta^g}(x)  &\displaystyle\mathop{\rightarrow}^\mathcal{P} & \sum_x K(\delta_g^g(x)) 
 f_{\theta^g}^{1+\alpha}(x) u_{jk\theta^g}(x), \label{EQ:47}  \\
 \sum_x K(\delta_n^g(x)) f_{\theta^g}^{1+\alpha}(x) u_{j\theta^g}(x) u_{k\theta^g}(x) &\displaystyle\mathop{\rightarrow}^\mathcal{P}& 
  \sum_x K(\delta_g^g(x)) f_{\theta^g}^{1+\alpha}(x) u_{j\theta^g}(x) u_{k\theta^g}(x). \nonumber \\ \label{EQ:48}     
\end{eqnarray}
Thus, combining (\ref{EQ:45}), (\ref{EQ:47}) and (\ref{EQ:48}), we get that,
\begin{eqnarray}
    \nabla_k \left( \sum_x K(\delta_n(x))f_{\theta}^{1+\alpha}(x)u_{j\theta}(x)|_{\theta=\theta^g} \right) 
\mathop{\rightarrow}^\mathcal{P} -J_g^{j,k}.
\end{eqnarray}
But we have,
\begin{eqnarray}
    2S_2 &=& (1+\alpha) \sum_{j,k} \left\{ \nabla_k \left( \sum_x K(\delta_n(x)) 
f_{\theta}^{1+\alpha}(x) u_{j\theta}(x)|_{\theta=\theta^g} \right) -  ( - J_g^{j,k})\right\} \nonumber \\
&& ~~~~~~~~~~~~~~~~~~~~~~~~~~~~~~~~~~~ \times  
(\theta_j - \theta_j^g)(\theta_k - \theta_k^g) \nonumber \\
 & &  ~~~~ + \sum_{j,k} \left\{ - \left((1+\alpha) J_g^{j,k} \right) 
(\theta_j - \theta_j^g)(\theta_k - \theta_k^g) \right\}. \label{EQ:50} \\ \nonumber
\end{eqnarray}
	Now the absolute value of the first term in above (\ref{EQ:50}) is $ < p^2 a^3$ with probability 
tending to one. And, the second term in (\ref{EQ:50}) is a negative definite quadratic form in the 
variables $(\theta_j - \theta_j^g)$. Letting $\lambda_1$ be the largest eigenvalue of 
$\frac{(1+\alpha)}{A} J_g$, the quadratic form is $ < \lambda_1 a^2$. Combining the two terms, 
we see that there exists $c > 0$ and $a_0 > 0$ such that for $a < a_0$, we have  $ S_2 < -c a^2$ 
with probability tending to one.\\

	Finally, considering the cubic term $S_3$, we have
\begin{eqnarray}
& &\nabla_{jkl} S_{(\alpha, \lambda)}(r_n, f_\theta)|_{\theta=\theta^*} = \nabla_{kl} 
\left( -(1+\alpha) ~ \sum_x K(\delta_n(x))f_{\theta}^{1+\alpha}(x)u_{j\theta}(x)|_{\theta=\theta^*} \right)
\nonumber \\
 && ~~~~~~~~~ =  -(1+\alpha)  \nabla_{l} \left(-\sum_x K'(\delta_n(x))\delta_n(x)f_{\theta}^{1+\alpha}(x)
 u_{j\theta}(x) u_{k\theta}(x)  \right. \nonumber \\
 && ~~~~~~~~~~~~~~~~~~~~~~~~~~~~~~~ \left. + \sum_x K(\delta_n(x)) f_{\theta}^{\alpha}(x)\nabla_{jk}f_\theta(x) 
\right) |_{\theta=\theta^*},  
\end{eqnarray}
or,
\begin{eqnarray}
&& - \frac{A}{1+\alpha} \nabla_{jkl} S_{(\alpha, \lambda)}(r_n, f_\theta)|_{\theta=\theta^*} \nonumber \\
&&~~~~~ =  \sum_x K''(\delta_n^*(x))
 \delta_n^*(x)^2 f_{\theta^*}^{1+\alpha}(x) u_{j\theta^*}(x) u_{k\theta^*}(x) u_{l\theta^*}(x)  \nonumber \\
& & ~~~~~~ - \sum_x K'(\delta_n^*(x)) \delta_n^*(x) f_{\theta^*}^{1+\alpha}(x) u_{j\theta^*}(x) u_{k\theta^*}(x) 
u_{l\theta^*}(x) \nonumber \\
& & ~~~~~~ - \sum_x K'(\delta_n^*(x)) \delta_n^*(x) f_{\theta^*}^{1+\alpha}(x) u_{j\theta^*}(x) u_{kl\theta^*}(x) \nonumber \\
& & ~~~~~~ - \sum_x K'(\delta_n^*(x)) \delta_n^*(x) f_{\theta^*}^{1+\alpha}(x)  u_{k\theta^*}(x) u_{jl\theta^*}(x) \nonumber \\
& & ~~~~~~  - \sum_x K'(\delta_n^*(x)) \delta_n^*(x) f_{\theta^*}^{1+\alpha}(x) \frac{\nabla_{jk}f_{\theta^*(x)}}{f_{\theta^*(x)}} 
 u_{l\theta^*}(x)   \nonumber \\
 & & ~~~~~~ + \sum_x K(\delta_n^*(x)) f_{\theta^*}^{1+\alpha}(x) \frac{\nabla_{jkl}f_{\theta^*(x)}}{f_{\theta^*(x)}} \nonumber \\
 & & ~~~~~~  + \sum_x K(\delta_n^*(x)) f_{\theta^*}^{1+\alpha}(x) \frac{\nabla_{jk}f_{\theta^*(x)}}{f_{\theta^*(x)}}  u_{l\theta^*}(x),   \label{EQ:52} 
\end{eqnarray}
where $\delta_n^*(x) = \frac{r_n(x)}{f_{\theta^*}(x)}$.	
We will now show that all the terms in the RHS of above (\ref{EQ:52}) are bounded.
Let us name the terms by (i) to (vii) respectively.
For the first term (i), we use (\ref{EQ:46}) to get
\begin{eqnarray}
   &&  |\sum_x K''(\delta_n^*(x)) \delta_n^*(x)^2 f_{\theta^*}^{1+\alpha}(x) u_{j\theta^*}(x) u_{k\theta^*}(x) u_{l\theta^*}(x)|  \nonumber \\ 
   && ~~~~ \le C_2 ~ \sum_x |\delta_n^*(x)| M_{j,k,l}(x)f_{\theta^*}(x) \nonumber \\
 & & ~~~~ = C_2 ~ \sum_x r_n^*(x) M_{j,k,l}(x) ~~~~ \mbox{(by CLT)} \nonumber \\
 &&~~~~ \rightarrow   C_2 ~ E_g[ M_{j,k,l}(X)]  < \infty.  ~~~~  \mbox{[by assumption (A6)]}
\end{eqnarray}
Thus term (i) is bounded. Now for the second term (ii), we again use (\ref{EQ:39}) to get 
\begin{eqnarray}
 &&    |\sum_x K'(\delta_n^*(x)) \delta_n^*(x) f_{\theta^*}^{1+\alpha}(x) u_{j\theta^*}(x) 
u_{k\theta^*}(x) u_{l\theta^*}(x)| \nonumber \\ 
&& ~~~~ \le C_1 ~ \sum_x |\delta_n^*(x)| M_{j,k,l}(x)f_{\theta^*}(x) \nonumber \\
 && ~~~~ = C_1 ~ \sum_x r_n^*(x) M_{j,k,l}(x) ~~~~ \mbox{(by CLT)} \nonumber \\
 && ~~~~ \rightarrow   C_1 ~ E_g[ M_{j,k,l}(X)]  < \infty,  ~~~~  \mbox{[by assumption (A6)]}
\end{eqnarray}
so that term (ii) is also bounded. Similarly the terms (iii), (iv) and (v) are bounded as in 
case of term (ii) and using (\ref{EQ:39}) and assumption (SA6). Next for the term (vi), we will consider the following:
\begin{eqnarray}
    |K(\delta)| = |\int_0^\delta K'(\delta) d\delta| \le C_1 |\delta|,
\end{eqnarray}
so that 
\begin{eqnarray}
    |K(\delta_n^*(x))| \le C_1 \frac{r_n(x)}{f_{\theta^*}(x)}.
\end{eqnarray}
Also,
\begin{eqnarray}
    \frac{\nabla_{jkl} f_{\theta^*}(x)}{f_{\theta^*}(x)} &=& u_{jkl\theta^*}(x) + u_{jk\theta^*}(x)
u_{l\theta^*}(x) + u_{jl\theta^*}(x)u_{k\theta^*}(x) \nonumber \\
 & &  + u_{j\theta^*}(x)u_{kl\theta^*}(x) + u_{j\theta^*}(x)u_{k\theta^*}(x)u_{l\theta^*}(x).
\end{eqnarray}
	So,
\begin{eqnarray}
     |\sum_x K(\delta_n^*(x)) f_{\theta^*}^{1+\alpha}(x) \frac{\nabla_{jkl} f_{\theta^*}(x)}
{f_{\theta^*}(x)}| &\le & C_1 ~ \sum_x r_n(x)  f_{\theta^*}^{\alpha}(x) |\frac{\nabla_{jkl} 
f_{\theta^*}(x)}{f_{\theta^*}(x)}| \nonumber \\
 & =&  C_1 ~ \sum_x r_n^*(x) M(x) ~~~~ \mbox{(by CLT)} \nonumber \\
 & \rightarrow &  C_1 ~ E_g[ M(X)]  < \infty  ~~~~  \nonumber \\ 
&& ~~~~~~~ \mbox{[by assumption (SA6)]}, 
\end{eqnarray}
where $M(x) = M_{jkl}(x) + M_{jk,l}(x) + M_{jl,k}(x) + M_{j,kl}(x) + M_{j,k,l}(x) $. 
Thus the term (vi) is bounded and also similarly the term (vii) is bounded.	
Hence, we have $|S_3|< b a^3$ on the sphere $Q_a$ with probability tending to one. 
Combining the three inequalities we get that 
$$
\max(S_1 + S_2 + S_3 ) < - c a^2 + (b + p) a^3,   
$$
which is strictly negative for $ a < \frac{c}{b+p}$.
Thus, for any sufficiently small $a$, there exists a sequence of roots $\theta_n = \theta_n(a)$ 
to the minimum $S$-divergence  estimating equation such that $P(||\theta_n - \theta^g||_2 < a)$ converges 
to one, where $||.||_2$ denotes the $L_2-$norm.

	It remains to show that we can determine such a sequence independent of $a$. 
For let $\theta_n^*$ be the root which is closes to $\theta^g$. This exists because 
the limit of a sequence of roots is again a root by the continuity of the $S$-divergence. 
This completes the proof of  the consistency part.
\\

\noindent
\textbf{Proof of the asymptotic Normality:}		For the Asymptotic normality, we expand 
$$\sum_x K(\delta_n(x)) f_{\theta}^{1+\alpha}(x)u_{\theta}(x)$$
 in Taylor series about $\theta = \theta^g$ to get
\begin{eqnarray}
    & & \sum_x K(\delta_n(x)) f_{\theta}^{1+\alpha}(x)u_{\theta}(x)  \nonumber \\
&& ~~~ = \sum_x K(\delta_n^g(x)) f_{\theta^g}^{1+\alpha}(x)u_{\theta^g}(x) \nonumber \\
 & & ~~~~~~~~~  + \sum_k (\theta_k - \theta_k^g) \nabla_k \left( \sum_x K(\delta_n(x))
 f_{\theta}^{1+\alpha}(x)u_{\theta}(x) \right)|_{\theta = \theta^g} \nonumber \\
 & & ~~~~~~~~~ + \frac{1}{2} \sum_{k,l} (\theta_k - \theta_k^g)(\theta_l - \theta_l^g)\nabla_{kl} 
\left(\sum_x K(\delta_n(x)) f_{\theta}^{1+\alpha}(x)u_{\theta}(x)\right)|_{\theta = \theta '},  \nonumber 
\\  \label{EQ:59}
\end{eqnarray}
where, $\theta '$ lies in between $\theta$ and $\theta^g$.\\

	Now, let $\theta_n$ be the solution of the minimum $S$-divergence estimating equation, which can be 
assumed to be consistent by the previous part. Replace $\theta$ by $\theta_n$ in above (\ref{EQ:59}) so that 
the LHS of the equation becomes zero and hence we get
\begin{eqnarray}
 && - \sqrt n \sum_x K(\delta_n^g(x)) f_{\theta^g}^{1+\alpha}(x)u_{\theta^g}(x)  \nonumber \\ 
  && ~~~ = 
 \sqrt n \sum_k (\theta_nk - \theta_k^g) \times \left\{ \nabla_k \left( \sum_x K(\delta_n(x)) f_{\theta}^{1+\alpha}(x)u_{\theta}(x) 
\right)|_{\theta = \theta^g} \right. \nonumber \\
& &~~~~~~ + \left. \frac{1}{2} \sum_{l} (\theta_nl - \theta_l^g)\nabla_{kl} 
\left(\sum_x K(\delta_n(x)) f_{\theta}^{1+\alpha}(x)u_{\theta}(x)\right)|_{\theta = \theta '}  \right\}. \label{EQ:60}
\end{eqnarray}
	Note that, the first term within the bracketed quantity in the RHS of above (\ref{EQ:60}) 
converges to $J_g$ with probability tending to one, while the second bracketed term is 
an $o_p(1)$ term (as proved in the proof of consistency part). 
Also, by using the Lemma \ref{LEM:lemma_3.4}, we get that
\begin{eqnarray}
   & & \sqrt n \sum_x K(\delta_n^g(x)) f_{\theta^g}^{1+\alpha}(x)u_{\theta^g}(x) \nonumber \\
&&  ~~~~~~~~~~~  =  \sqrt n \sum_x  [K(\delta_n^g(x)) - K(\delta_g^g(x))]  
f_{\theta^g}^{1+\alpha}(x)u_{\theta^g}(x) \nonumber \\
&& ~~~~~~~~~~~  =  S_{1n}|_{\theta=\theta^g} \mathop{\rightarrow}^\mathcal{D} N_p(0, V_g). 
\end{eqnarray}

	Therefore, by Lehmann (1983, Lemma 4.1), it follows that $\sqrt n (\theta_n - \theta^g)$
 has asymptotic distribution as $N_p( 0, J_g^{-1} V_g J_g^{-1} )$.
\hfill{$\square$} 

\bigskip

\begin{corollary}
	When the true distribution $G$ belongs to the model family, i.e., $G = F_\theta$ for 
some $\theta \in \Theta$, then $\sqrt n (\theta_n - \theta)$ has asymptotic distribution 
as $N_p( 0, J_\alpha^{-1} V_\alpha J_\alpha^{-1} )$, where 
\begin{eqnarray}
     J_\alpha = J_\alpha(f_\theta) &=&  E_{f_\theta}[u_\theta(X)u_\theta(X)^T f_\theta^{\alpha}(X)] = 
\int u_\theta(x)u_\theta^T(x)f_\theta^{1+\alpha}(x) dx, \\
	 V_\alpha = V_\alpha(f_\theta) &=& V_{f_\theta}[u_\theta(X)f_\theta^{\alpha}(X)] = 
\int u_\theta(x)u_\theta^T(x)f_\theta^{1+2\alpha}(x) dx - \xi \xi^T, \\
     \xi = \xi_\alpha(f_\theta) &=& E_{f_\theta}[u_\theta(X)f_\theta^{\alpha}(X)] = 
\int u_\theta(x)f_\theta^{1+\alpha}(x)dx.
\end{eqnarray}
Note that, this asymptotic distribution is independent of the parameter 
$\lambda$ in the $S$-divergence Family.\\
\end{corollary}
\noindent
\textbf{Proof:}	Note that, under  $G = F_\theta$ for some $\theta \in \Theta$, 
we get $\delta_g^g(x) = 1 ~~~ \forall x$ so that $K(\delta_g^g(x))= 0$ 
and $K'(\delta_g^g(x))= A $. Thus the result follows from the above theorem
by noting that $J_g=J_\alpha$ and $V_g=V_\alpha$.
\hfill{$\square$}

\bigskip
Note that, the asymptotic variance of the proposed minimum $S$-divergence estimator
depends only on the parameter $\alpha$ at the model family and hence coincides with that 
of the minimum density power divergence estimator (which corresponds to the case $\lambda=0$) of Basu et al.~(1998).
Further, interestingly, at the case $\alpha=0$, this asymptotic variance of the MSDE
coincides with the inverse of the Fisher information matrix $I(\theta) = E[u_\theta^Tu_\theta]$ 
($J_0=I(\theta)$ and $V_0=I(\theta)$) irrespectively of $\lambda$ as expected;
note that the MSDE with $\alpha=\lambda=0$ is in fact the MLE having the minimum asymptotic variance 
at the model. Thus, the asymptotic relative efficiency (ARE) of the minimum $S$-divergence estimators 
$\hat{\theta}_{(\alpha, \lambda)}$ can be calculated by comparing its asymptotic variance 
with that under the case $\alpha=\lambda=0$. For example, when $\theta \in \mathbb{R}$, we can define
$$
{\rm ARE}\left(\hat{\theta}_{(\alpha, \lambda)}\right) 
= \frac{J_0^{-1} V_0 J_0^{-1}}{J_\alpha^{-1} V_\alpha J_\alpha^{-1}} 
= \frac{I^{-1}}{J_\alpha^{-1} V_\alpha J_\alpha^{-1}}.
$$
This measure is easy to calculate for the common parametric models.
However, it is to be noted that the all the above asymptotic results hold under the assumptions 
(SA1)--(SA7) and one should check these conditions before applying the proposed MSDE to any parametric model.
We have verified this conditions to hold for most common parametric models.
In the following we will present the examples of two particular models to illustrate the validity of the assumptions 
(SA1)--(SA7) and the usefulness of the asymptotic results derived above.

\bigskip
\noindent
\textbf{Example 1: Poisson Model}\\
First let us consider the popular parametric model of Poisson distribution with mean $\theta$.
We will verify that conditions (SA1)--(SA7) hold for this model assuming that the 
 true density $g$ is also a Poisson distribution with mean $\theta^g=\theta_0$.
Clearly, the Poisson model family is identifiable with the open parameter space $\Theta=(0,\infty)$
and it has support $\chi = \{0, 1, 2, 3, \ldots\}$, the set of all non-negative integers,
which is independent of the mean parameter $\theta$. 
Further, the density of the Poisson distribution  is continuous in $\theta$ and is given by 
$$
f_\theta(x) = \frac{\theta^x}{x!}e^{-\theta}, ~~~~~ x \in \chi.
$$
Thus, clearly (SA1)--(SA3) holds for this model family. 
Next, note that, in this case, we have
$$
\nabla \log f_\theta(x) = \frac{x}{\theta} - 1, ~~ 
\nabla^2 \log f_\theta(x) = -\frac{x}{\theta^2}, ~~ 
\nabla^3 \log f_\theta(x) = \frac{2x}{\theta^3}, ~~~~~ x \in \chi.
$$
So, using the boundedness of the functions $\frac{1}{z^p}e^{-z}$, 
where $p$ is a positive integer, on the domain $z\in (0, \infty)$,
one can easily show that the conditions (SA5) and (SA6) hold true.
Further, using the above forms, we have, for the Poisson model,
$$
J_\alpha = \sum_{x=0}^{\infty}~ \left(\frac{x}{\theta} - 1\right)^2 
\frac{\theta^{(1+\alpha)x}}{(x!)^{(1+\alpha)}}e^{-(1+\alpha)\theta},
$$ 
which is clearly a positive real number implying (SA4) holds.
Finally, to show (SA7), note that
$$
\left( \frac{g(x)}{f_{\theta}(x)} \right)^{A-1} = \left( \frac{f_{\theta_0}(x)}{f_{\theta}(x)} \right)^{A-1} 
= \left(\frac{\theta_0}{\theta}\right)^{x(A-1)} e^{-(A-1)(\theta_0-\theta)},
$$
which is clearly uniformly bounded in $\theta \in (0, \infty)$. 
Hence all the assumptions (SA1)--(SA7) hold under the Poisson model.

Now, applying the above Theorem 1, the MSDE of the Poisson parameter $\theta$ 
is consistent and asymptotically normal with variance given by
$J_\alpha^{-1} V_\alpha J_\alpha^{-1}$, where $J_\alpha$ is as defined above and 
$$
V_\alpha = \sum_{x=0}^{\infty}~ \left(\frac{x}{\theta} - 1\right)^2 
\frac{\theta^{(1+2\alpha)x}}{(x!)^{(1+2\alpha)}}e^{-(1+2\alpha)\theta}
- \left(\sum_{x=0}^{\infty}~ \left(\frac{x}{\theta} - 1\right) 
\frac{\theta^{(1+\alpha)x}}{(x!)^{(1+\alpha)}}e^{-(1+\alpha)\theta}\right)^2.
$$
This asymptotic variance can be calculated by a simple numerical summation
and can be compared with the corresponding Fisher information matrix $I(\theta)$ to 
examine the asymptotic relative efficiencies of the MSDEs.
Note that, for the Poisson model, the Fisher information matrix is given by 
$$
I(\theta) = \sum_{x=0}^{\infty}~ \left(\frac{x}{\theta} - 1\right)^2 
\frac{\theta^{x}}{x!}e^{-\theta} = \frac{1}{\theta}.
$$
Table \ref{TAB:ARE_MSDE} presents the value of ARE for several MSDEs; note that 
as seen in Corollary 1 the asymptotic variance and hence the ARE of the MSDE is independent of the 
parameter $\lambda$. It can be seen from the table that the ARE of MSDE is maximum ($100\%$)
if $\alpha=0$; as $\alpha$ increases the efficiency decreases.

 \begin{table}
\caption{The asymptotic relative efficiency of the MSDE under Poisson($\theta$) and Geometric($\theta$) Model for different values of $\alpha$ and $\theta$}
		\begin{tabular}{l r|  rrrrrrr} 
				\hline\noalign{\smallskip}
\multicolumn{1}{l}{Model} & \multicolumn{1}{c|}{$\theta$} & \multicolumn{7}{c}{$\alpha$}\\
\noalign{\smallskip}\hline\noalign{\smallskip}
	 & 	&	0	&	0.05	&	0.1	&	0.3 &	0.5	&	0.7	& 1	\\ 
\noalign{\smallskip}\hline\noalign{\smallskip}
	&	2	&	100	&	99.62	&	98.77	&	93.06	&	86.15	&	79.55	&	71.17	\\
	&	3	&	100	&	99.66	&	98.82	&	92.86	&	85.18	&	77.42	&	68.22	\\
Poisson	&	5	&	100	&	99.61	&	98.80	&	92.38	&	84.19	&	76.96	&	66.47	\\
	&	10	&	100	&	99.66	&	98.75	&	92.07	&	83.86	&	76.07	&	65.69	\\
	&	15	&	100	&	99.66	&	98.83	&	92.09	&	83.76	&	75.71	&	65.59	\\
	\noalign{\smallskip}\hline\noalign{\smallskip}
	&	0.1	&	100	&	99.10	&	96.78	&	81.93	&	68.42	&	59.24	&	51.06	\\
	&	0.2	&	100	&	99.10	&	96.79	&	82.01	&	68.59	&	59.49	&	51.45	\\
Geometric	&	0.5	&	100	&	99.14	&	96.92	&	82.90	&	70.37	&	62.19	&	55.64	\\
	&	0.7	&	100	&	99.21	&	97.19	&	84.71	&	73.98	&	67.54	&	63.61	\\
	&	0.9	&	100	&	99.43	&	98.03	&	90.04	&	84.07	&	81.56	&	82.15	\\
\noalign{\smallskip}\hline
		\end{tabular}
\label{TAB:ARE_MSDE}
\end{table}

\bigskip
\noindent
\textbf{Example 2: Geometric Model}\\
Now consider another popular parametric model family of Geometric distribution with success probability $\theta$.
Again we will verify conditions (SA1)--(SA7) assuming that the true density $g$ 
belongs to the model family with parameter value $\theta^g=\theta_0$.
Clearly, the Geometric family is identifiable with the open parameter space $\Theta=(0,1)$
and support $\chi = \{1, 2, 3, \ldots\}$, the set of all positive integers,
which is independent of the parameter $\theta$. 
The Geometric model also has a continuous  density (in $\theta$) given by 
$$
f_\theta(x) = \theta (1-\theta)^{x-1}, ~~~~~ x \in \chi.
$$
So, assumptions (SA1)--(SA3) holds for the Geometric model family. 
Also, in this case, we have
$$
\nabla \log f_\theta(x) = \frac{1}{\theta} - \frac{x-1}{1-\theta}, ~~ 
\nabla^2 \log f_\theta(x) = -\frac{1}{\theta^2} - \frac{x-1}{(1-\theta)^2}, ~~
$$
$$ 
\nabla^3 \log f_\theta(x) = \frac{2}{\theta^3} - \frac{2(x-1)}{(1-\theta)^3}, ~~~~~ x \in \chi.
$$
Thus, one can easily prove the conditions (SA5) and (SA6) hold true for this Geometric model.
Further, we get
$$
J_\alpha = \frac{\theta^{(\alpha-1)}[t_\alpha(\theta)^2 - \theta(\theta+2)t_\alpha(\theta) + 2\theta^2]}{
(1-\theta)^2 t_\alpha(\theta)},
$$ 
where $t_\alpha(\theta)=\left(1-(1-\theta)^{(1+\alpha)}\right)$. 
Clearly, $J_\alpha$ is a positive real number for all $\theta \in (0, 1)$ and so (SA4) holds.
Finally, we have
$$
\left( \frac{g(x)}{f_{\theta}(x)} \right)^{A-1} = \left( \frac{f_{\theta_0}(x)}{f_{\theta}(x)} \right)^{A-1} 
= \left(\frac{\theta_0}{\theta}\right)^{x(A-1)},
$$
which is clearly uniformly bounded in $\theta \in (0, 1)$ by $C=1$. 
Hence all the assumptions (SA1)--(SA7) holds under the Geometric model also.

Now, applying Theorem 1, we have that the MSDE of parameter $\theta$ 
is consistent and asymptotically normal with variance 
$J_\alpha^{-1} V_\alpha J_\alpha^{-1}$, where 
$$
V_\alpha = \frac{\theta^{(2\alpha-1)}[t_{2\alpha}(\theta)^2 - \theta(\theta+2)t_{2\alpha}(\theta) + 2\theta^2]}{
(1-\theta)^2 t_{2\alpha}(\theta)} - \left(\frac{\theta^{2\alpha}(1-(1-\theta)^\alpha)^2}{t_\alpha(\theta)^4}\right).
$$
Under the Geometric model, the Fisher information matrix $I(\theta)$ has the simple form  
$$
I(\theta) = \frac{1}{\theta^2(1-\theta)}.
$$
We can again compute the ARE of the MSDEs of the Geometric parameter using the above expressions,
which is reported in Table \ref{TAB:ARE_MSDE}.
Clearly, the table shows that the asymptotic relative efficiency decreases as $\alpha$ increases.
However, there is no significant loss in efficiency at the smaller positive values of $\alpha$.

\section{Real Data Examples}\label{SEC:example}


\subsection{Drosophila Data: Poisson Model}

	Here we consider a chemical mutagenicity experiment. 
These data were analyzed previously by Simpson (1987). The details of the experimental protocol 
are available in Woodruff et al. (1984). In a sex linked recessive lethal test in Drosophila (fruit flies), 
the experimenter exposed groups of male flies to different doses of a chemical to be screened. 
Each male was then mated with unexposed females. Sampling 100 daughter flies from each male 
(roughly), the number of daughters carrying a recessive lethal mutation on the $X$ 
chromosome was noted. The data set consisted of the observed frequencies
of males having $0, 1, 2, \cdots$ recessive lethal daughters. For our purpose, 
we consider two specific experimental runs --- one on the day 28 and second on day 177. 
The data of the first run consist of two small outliers with observed frequencies $d = (23, 3, 1, 1)$ 
at $x = (0,1,3,4)$ and that of second run consists 
of observed frequencies $d = (23, 7, 3, 1)$ at $x = (0,1,2,91)$ with a large outlier at $91$.

 \begin{table}
	\caption{The estimate of the Poisson parameter for different values of $\alpha$ and $\lambda$ for Drosophila data without outlier: First Experimental Run}
		\begin{tabular}{r| l l l l l l l  l} 
				\hline\noalign{\smallskip}
		 $\lambda$	&	$\alpha = $ 0	&	$\alpha = $ 0.1	&	$\alpha = $ 0.25	&	$\alpha = $ 0.4	&	$\alpha = $ 0.5	&	$\alpha = $ 0.6	&	$\alpha = $ 0.8	&	$\alpha = $ 1	\\ \noalign{\smallskip}\hline\noalign{\smallskip}
$-1$	&	--	&	0.08	&	0.11	&	0.12	&	0.12	&	0.12	&	0.13	&	0.13	\\
$-0.7$	&	0.09	&	0.10	&	0.12	&	0.12	&	0.12	&	0.13	&	0.13	&	0.13	\\
$-0.5$	&	0.10	&	0.11	&	0.12	&	0.12	&	0.12	&	0.13	&	0.13	&	0.13	\\
$-0.3$	&	0.11	&	0.12	&	0.12	&	0.12	&	0.12	&	0.13	&	0.13	&	0.13	\\
$-0.1$	&	0.11	&	0.12	&	0.12	&	0.12	&	0.13	&	0.13	&	0.13	&	0.13	\\
0	&	0.12	&	0.12	&	0.12	&	0.12	&	0.13	&	0.13	&	0.13	&	0.13	\\
0.5	&	0.12	&	0.12	&	0.12	&	0.13	&	0.13	&	0.13	&	0.13	&	0.13	\\
1	&	0.12	&	0.12	&	0.13	&	0.13	&	0.13	&	0.13	&	0.13	&	0.13	\\
1.3	&	0.12	&	0.12	&	0.13	&	0.13	&	0.13	&	0.13	&	0.13	&	0.13	\\
1.5	&	0.12	&	0.12	&	0.13	&	0.13	&	0.13	&	0.13	&	0.13	&	0.13	\\
2	&	0.12	&	0.13	&	0.13	&	0.13	&	0.13	&	0.13	&	0.13	&	0.13	\\
\noalign{\smallskip}\hline
		\end{tabular}
\label{TAB:WO_D1}
\end{table}

 \begin{table}
	\caption{The estimate of the Poisson parameter for different values of $\alpha$ and $\lambda$ for Drosophila data with outlier: First Experimental Run}
		\begin{tabular}{r| l l l l l l l  l} 
				\hline\noalign{\smallskip}
		 $\lambda$	&	$\alpha = $ 0	&	$\alpha = $ 0.1	&	$\alpha = $ 0.25	&	$\alpha = $ 0.4	&	$\alpha = $ 0.5	&	$\alpha = $ 0.6	&	$\alpha = $ 0.8	&	$\alpha = $ 1	\\ \noalign{\smallskip}\hline\noalign{\smallskip}
$-1$	&	--	&	0.08	&	0.11	&	0.13	&	0.14	&	0.14	&	0.15	&	0.16	\\
$-0.7$	&	0.10	&	0.11	&	0.13	&	0.14	&	0.14	&	0.15	&	0.16	&	0.16	\\
$-0.5$	&	0.13	&	0.13	&	0.13	&	0.14	&	0.14	&	0.15	&	0.16	&	0.16	\\
$-0.3$	&	0.18	&	0.15	&	0.14	&	0.14	&	0.14	&	0.15	&	0.16	&	0.16	\\
$-0.1$	&	0.29	&	0.22	&	0.16	&	0.15	&	0.15	&	0.15	&	0.16	&	0.16	\\
0	&	0.36	&	0.26	&	0.18	&	0.15	&	0.15	&	0.15	&	0.16	&	0.16	\\
0.5	&	0.59	&	0.49	&	0.34	&	0.21	&	0.17	&	0.16	&	0.16	&	0.16	\\
1	&	0.70	&	0.63	&	0.49	&	0.32	&	0.18	&	0.17	&	0.16	&	0.16	\\
1.3	&	0.75	&	0.68	&	0.55	&	0.39	&	0.28	&	0.19	&	0.16	&	0.16	\\
1.5	&	0.77	&	0.71	&	0.59	&	0.44	&	0.32	&	0.25	&	0.16	&	0.16	\\
2	&	0.81	&	0.76	&	0.66	&	0.52	&	0.40	&	0.27	&	0.16	&	0.16	\\
\noalign{\smallskip}\hline
		\end{tabular}
\label{TAB:W_D1}
\end{table}

	Poisson models are fitted to the data for this experimental runs by estimating 
the Poisson parameter using minimum $S$-divergence  estimation for several values of $\alpha$ and 
$\lambda$. A quick look at the observed frequencies for the experimental run reveals that 
there is an exceptionally large count -- where one male is reported to have produced 91 
daughters with the recessive lethal mutation. We estimate the Poisson parameter from this 
data with the outlying observation and without that outlying observation. The difference 
in these two estimates gives an indication of the robust behavior (or lack thereof) of different 
Minimum $S$-divergence estimators. Our findings are reported in Tables \ref{TAB:WO_D1} to \ref{TAB:W_D2}.

The values of the minimum $S$-divergence estimators given in these tables clearly demonstrate their 
robustness with respect to the outlying value for all $\alpha \in [0,1]$ if $\lambda < 0$. 
For $\lambda = 0$ the minimum $S$-divergence estimators are also robust for large values of $\alpha$, 
but smaller values of $\alpha$ are highly non-robust (note that $\alpha = 0$ and $ \lambda = 0$ 
gives the MLE).  For $\lambda > 0$ the MSDEs corresponding to small values of $\alpha$ 
close to zero are highly sensitive to the outlier; this sensitivity decreases with $\alpha$, 
and eventually the outlier has negligible effect on the estimator when $\alpha$ is very close to 1. 
The robustness of the estimators decrease sharply with increasing $\lambda$ except when $\alpha = 1$;
note that this particular case with $\alpha=1$ gives the $L_2$ divergence irrespective of the value of $\lambda$
which is highly robust but inefficient.

 \begin{table}
	\caption{The estimate of the Poisson parameter for different values of $\alpha$ and $\lambda$ for Drosophila data without outlier: Second Experimental Run}
		\begin{tabular}{r| l l l l l l l  l} 
				\hline\noalign{\smallskip}
		 $\lambda$	&	$\alpha = $ 0	&	$\alpha = $ 0.1	&	$\alpha = $ 0.25	&	$\alpha = $ 0.4	&	$\alpha = $ 0.5	&	$\alpha = $ 0.6	&	$\alpha = $ 0.8	&	$\alpha = $ 1	\\ \noalign{\smallskip}\hline\noalign{\smallskip}
$-1$	&	--	&	0.29	&	0.35	&	0.36	&	0.36	&	0.35	&	0.35	&	0.35	\\
$-0.7$	&	0.34	&	0.35	&	0.36	&	0.36	&	0.36	&	0.36	&	0.35	&	0.35	\\
$-0.5$	&	0.36	&	0.37	&	0.37	&	0.36	&	0.36	&	0.36	&	0.35	&	0.35	\\
$-0.3$	&	0.38	&	0.38	&	0.37	&	0.37	&	0.36	&	0.36	&	0.35	&	0.35	\\
$-0.1$	&	0.39	&	0.39	&	0.38	&	0.37	&	0.37	&	0.36	&	0.35	&	0.35	\\
0	&	0.39	&	0.39	&	0.38	&	0.37	&	0.37	&	0.36	&	0.35	&	0.35	\\
0.5	&	0.41	&	0.40	&	0.39	&	0.38	&	0.37	&	0.36	&	0.35	&	0.35	\\
1	&	0.42	&	0.42	&	0.40	&	0.39	&	0.32	&	0.37	&	0.36	&	0.35	\\
1.3	&	0.43	&	0.42	&	0.41	&	0.39	&	0.38	&	0.37	&	0.36	&	0.35	\\
1.5	&	0.43	&	0.42	&	0.41	&	0.39	&	0.38	&	0.37	&	0.36	&	0.35	\\
2	&	0.44	&	0.43	&	0.42	&	0.40	&	0.39	&	0.37	&	0.36	&	0.35	\\
\noalign{\smallskip}\hline
		\end{tabular}
\label{TAB:WO_D2}
\end{table}

 \begin{table}
	\caption{The estimate of the Poisson parameter for different values of $\alpha$ and $\lambda$ for Drosophila data with outlier: Second Experimental Run}
		\begin{tabular}{r| l l l l l l l  l} 
		\hline\noalign{\smallskip}
 $\lambda$	&	$\alpha = $ 0	&	$\alpha = $ 0.1	&	$\alpha = $ 0.25	&	$\alpha = $ 0.4	&	$\alpha = $ 0.5	&	$\alpha = $ 0.6	&	$\alpha = $ 0.8	&	$\alpha = $ 1	\\ \noalign{\smallskip}\hline\noalign{\smallskip}
$-1$	&	--	&	0.30	&	0.35	&	0.36	&	0.36	&	0.36	&	0.36	&	0.36	\\
$-0.7$	&	0.34	&	0.36	&	0.37	&	0.37	&	0.37	&	0.37	&	0.36	&	0.36	\\
$-0.5$	&	0.36	&	0.37	&	0.37	&	0.37	&	0.37	&	0.37	&	0.37	&	0.36	\\
$-0.3$	&	0.38	&	0.38	&	0.38	&	0.37	&	0.37	&	0.37	&	0.37	&	0.36	\\
$-0.1$	&	0.39	&	0.39	&	0.38	&	0.38	&	0.37	&	0.37	&	0.37	&	0.36	\\
0	&	3.03	&	0.39	&	0.39	&	0.38	&	0.37	&	0.37	&	0.37	&	0.36	\\
0.5	&	31.31	&	30.28	&	25.12	&	0.39	&	0.38	&	0.37	&	0.37	&	0.36	\\
1	&	32.20	&	31.84	&	30.79	&	27.08	&	0.99	&	0.38	&	0.37	&	0.36	\\
1.3	&	32.40	&	32.15	&	31.48	&	29.71	&	24.93	&	0.38	&	0.37	&	0.36	\\
1.5	&	32.50	&	32.29	&	31.76	&	30.48	&	27.78	&	22.54	&	0.37	&	0.36	\\
2	&	33.22	&	32.50	&	32.15	&	31.43	&	30.28	&	26.24	&	0.37	&	0.36	\\
\noalign{\smallskip}\hline
		\end{tabular}
\label{TAB:W_D2}
\end{table}

\subsection{Peritonitis Incidence Data: Geometric Model}

Now we will consider another interesting real data example on the incidence of peritonitis 
for 390 kidney patients (Basu et al., 2011, Table 2.4).  
Basu et al.~(2011) examined this data by fitting a geometric distribution with parameter 
$\theta$ (success probability) around 0.5 and observed that there are two mild to moderate outliers 
at the points $10$ and $12$ that moderately affect the non-robust estimators. 
The effect of outliers is not so dramatic here as in the previous example due to its large sample size.
Thus, this data set  provides another interesting situation to examine the performance of any robust estimator.

We will apply the proposed minimum $S$-divergence estimators to estimate the geometric parameter 
for this data set --- once ignoring the two outlying observations and once considering the full data. 
The estimated values ae reported in Tables \ref{TAB:WO_PK} and \ref{TAB:W_PK} respectively. 
Again, we can see from the tables that the minimum $S$-divergence estimators differ significantly 
even in the presence of mild outliers for all smaller values of $\alpha$ with $\lambda > 0$;
but the MSDEs with $\lambda<0$ or larger values of $\alpha$ with $\lambda \geq 0$ 
remain more stable with respect to the outlying observations as seen in the previous example.

\begin{table}
\caption{The estimate of the Geometric parameter for different values of $\alpha$ and $\lambda$ for Peritonitis Incidence Data  without two outliers}
\resizebox{\textwidth}{!}{
\begin{tabular}{r| l l l l l l l  l} 
\hline\noalign{\smallskip}
$\lambda$	&	$\alpha = $ 0	&	$\alpha = $ 0.1	&	$\alpha = $ 0.25	&	$\alpha = $ 0.4	&	$\alpha = $ 0.5	&	$\alpha = $ 0.6	&	$\alpha = $ 0.8	&	$\alpha = $ 1	\\ \noalign{\smallskip}\hline\noalign{\smallskip}
$-1$	&	--	&	0.5392	&	0.5155	&	0.5099	&	0.5089	&	0.5088	&	0.5091	&	0.5097	\\
$-0.7$	&	0.5257	&	0.5170	&	0.5107	&	0.5085	&	0.5082	&	0.5087	&	0.5090	&	0.5097	\\
$-0.5$	&	0.5176	&	0.5128	&	0.5090	&	0.5079	&	0.5079	&	0.5086	&	0.5090	&	0.5097	\\
$-0.3$	&	0.5133	&	0.5101	&	0.5078	&	0.5074	&	0.5076	&	0.5085	&	0.5089	&	0.5097	\\
$-0.1$	&	0.5104	&	0.5082	&	0.5069	&	0.5069	&	0.5073	&	0.5084	&	0.5089	&	0.5097	\\
0	&	0.5092	&	0.5074	&	0.5064	&	0.5067	&	0.5072	&	0.5083	&	0.5089	&	0.5097	\\
0.5	&	0.5047	&	0.5042	&	0.5046	&	0.5057	&	0.5065	&	0.5081	&	0.5088	&	0.5097	\\
1	&	0.5014	&	0.5018	&	0.5030	&	0.5047	&	0.5059	&	0.5079	&	0.5087	&	0.5097	\\
1.3	&	0.4998	&	0.5005	&	0.5022	&	0.5042	&	0.5056	&	0.5078	&	0.5086	&	0.5097	\\
1.5	&	0.4987	&	0.4996	&	0.5016	&	0.5039	&	0.5053	&	0.5077	&	0.5085	&	0.5097	\\
2	&	0.4964	&	0.4977	&	0.5003	&	0.5031	&	0.5048	&	0.5075	&	0.5084	&	0.5097	\\
\noalign{\smallskip}\hline
\end{tabular}}
\label{TAB:WO_PK}
\end{table}
\begin{table}
\caption{The estimate of the Geometric parameter for different values of $\alpha$ and $\lambda$ for Peritonitis Incidence Data  with outlier}
\resizebox{\textwidth}{!}{
\begin{tabular}{r| l l l l l l l  l} 
\hline\noalign{\smallskip}
$\lambda$	&	$\alpha = $ 0	&	$\alpha = $ 0.1	&	$\alpha = $ 0.25	&	$\alpha = $ 0.4	&	$\alpha = $ 0.5	&	$\alpha = $ 0.6	&	$\alpha = $ 0.8	&	$\alpha = $ 1	\\ \noalign{\smallskip}\hline\noalign{\smallskip}
$-1$	&	--	&	0.5346	&	0.5134	&	0.5090	&	0.5084	&	0.5087	&	0.5090	&	0.5097	\\
$-0.7$	&	0.5193	&	0.5129	&	0.5087	&	0.5077	&	0.5078	&	0.5085	&	0.5090	&	0.5097	\\
$-0.5$	&	0.5104	&	0.5082	&	0.5068	&	0.5069	&	0.5074	&	0.5084	&	0.5089	&	0.5097	\\
$-0.3$	&	0.5044	&	0.5046	&	0.5053	&	0.5063	&	0.5070	&	0.5083	&	0.5088	&	0.5097	\\
$-0.1$	&	0.4990	&	0.5013	&	0.5038	&	0.5057	&	0.5066	&	0.5082	&	0.5088	&	0.5097	\\
0	&	0.4962	&	0.4996	&	0.5031	&	0.5053	&	0.5065	&	0.5082	&	0.5088	&	0.5097	\\
0.5	&	0.4798	&	0.4893	&	0.4986	&	0.5036	&	0.5055	&	0.5079	&	0.5087	&	0.5097	\\
1	&	0.4609	&	0.4751	&	0.4920	&	0.5012	&	0.5044	&	0.5076	&	0.5085	&	0.5097	\\
1.3	&	0.4503	&	0.4657	&	0.4866	&	0.4993	&	0.5035	&	0.5074	&	0.5085	&	0.5097	\\
1.5	&	0.4439	&	0.4595	&	0.4824	&	0.4978	&	0.5029	&	0.5073	&	0.5084	&	0.5097	\\
2	&	0.4304	&	0.4455	&	0.4708	&	0.4926	&	0.5007	&	0.5070	&	0.5083	&	0.5097	\\
\noalign{\smallskip}\hline
\end{tabular}}
\label{TAB:W_PK}
\end{table}

\section{Integration of Empirical and Asymptotic Results: Choice of tuning parameters}
\label{SEC:choice_par}

In the last two sections we have observed the following: 
(a) The asymptotic distributions of the proposed minimum $S$-divergence estimators 
(and hence their asymptotic relative efficiencies) are independent of the parameter $\lambda$; and 
(b) the behavior of the estimators with respect to robustness against outliers are widely different 
for different combinations of $\alpha$ and $\lambda$, and sometimes even vary greatly 
over different values of $\lambda$ for fixed $\alpha$. 
These observations indicate that a proper discussion of the role of the two tuning parameters 
are important in this context, and in this connection we record the following points. 
For part of this discussion we borrow from the Ghosh et al.~(2013) paper, 
which describes the robustness issues related to the $S$-divergence, unlike the present paper 
which primarily concentrates on the asymptotic efficiency results of the corresponding estimators. 

\begin{enumerate}

\item The influence function of the minimum $S$-divergence estimators are independent of $\lambda$. 
This has, in fact, been directly observed by Ghosh et al.~(2013) who evaluated the influence function 
of the minimum $S$-divergence estimators; see  Ghosh et al.~(2013), Section 4.2. 

\item Our examples clearly show, however, that the true stability of our proposed estimators 
against outliers are not identical over $\lambda$ for fixed values of $\alpha$. 
The estimators at $\alpha = 0$ (or low values close to zero) are highly influenced by 
the choice of the value of $\lambda$ under the presence of outliers. 

\item This indicates, further, that the influence function of the $S$-divergence estimators 
are not able to fully predict the robustness behavior of the minimum $S$-divergence estimators. 
Ghosh et al.~(2013), have, in fact, demonstrated that for different choices of the tuning parameters, 
the second order influence function prediction can be widely different from the first, 
and the discrepancy may be in either direction. We refer the reader to Ghosh et al.~for 
an extensive discussion of this phenomenon, including theoretical calculations of the first 
and second order influence functions, extensive simulations and detailed graphical studies.  
\end{enumerate}

Another issue of importance that immediately presents itself on the basis of the above discussion 
is the choice of the tuning parameters which could be the most appropriate in a particular situation, 
where the experimenter is unaware of the purity of the data or about the nature of possible contaminations. 
This is clearly an issue which will require more research. 
However, on the basis of our empirical findings of Section \ref{SEC:example} and 
theoretical efficiencies of Section \ref{SEC:asymptotics} (Table \ref{TAB:ARE_MSDE}), 
it would appear that low values of $\alpha$ 
(say between 0.1 and 0.25) with moderately large negative values of $\lambda$ 
(say beteen $-0.3$ and $-0.5$) should be the more appropriate choices. 

A further obvious application of the divergences considered in this paper would be in the case of 
testing of parametric hypothesis. Some indications of the potential of the proposed divergence in this 
connection has been presented in Ghosh et al.~(2014), where one uses the form of the $S$-divergence 
to quantify the discrepancy between the null distribution and the empirical distribution. 
A simplifying application is to use the minimum DPD estimator in this connection in place of 
the actual minimum $S$-divergence estimator. As the distribution of our proposed estimators 
in Section \ref{SEC:asymptotics} do not depend on $\lambda$, 
the test statistics proposed by Ghosh et al.~(2014) have the same asymptotic distributions 
as one would get if the original minimum $S$-divergence estimators were used.

The proposed $S$-divergence based inference has enough potential for application  in 
several applied field where the observed data is supposed to contains outlying observations.
One possible application of the minimum $S$-divergence estimator in the context of reliability 
is described in Ghosh, Maji and Basu (2013). 
Further works are to be done in future to examine its properties in other applications.

\section{Conclusion}
\label{SEC:conclusion}
The $S$-divergence family generates a large class of divergence measures having several important properties. 
Thus, the minimum divergence estimators obtained by minimizing these different members of 
the $S$-divergences family  also have several interesting properties in terms of their efficiency and robustness. 
In this present paper, we have proved the asymptotic properties of the minimum $S$-divergence estimators 
under the discrete set-up. Interestingly, we have seen that the asymptotic distributions of 
the minimum $S$-divergence estimators at the model is independent of one defining parameter $\lambda$, 
although their robustness depends on this parameter value. 
Indeed, considering the minimum $S$-divergence estimators as members of a grid constructed based on 
its defining parameters $\lambda$ and $\alpha$, we can clearly observe a triangular region of 
non-robust estimators corresponding to the large $\lambda $ and small $\alpha$ values and 
a region of highly robust estimators corresponding to moderate $\alpha$ and large negative $\lambda$ values.
As a future work, we need to prove all the properties of the minimum $S$-divergence estimators 
under the continuous models. However, under the continuous model, we need to use the kernel smoothing 
to estimate the true density $g$ and hence proving the asymptotic properties will inherit 
all the complications of the kernel estimation like bandwidth selection etc. 
We will try to solve these issues in our subsequent papers.

\begin{acknowledgements}
The author would like to thank his Ph.D.~supervisor Prof. Ayanendranath Basu (Indian Statistical Institute, India) 
for his sincere guidance and valuable comments about this work. 
The author also wishes to thank two anonymous referees for their comments 
that helped to improve the content and presentation of this paper.
\end{acknowledgements}


\end{document}